\documentclass{webofc}
\usepackage[varg]{txfonts}   

\usepackage{lineno}
\usepackage{hyperref}

\newcommand{\W}        {\ensuremath{\rm{W}^\pm}\xspace}
\newcommand{\Wminus}   {\ensuremath{\rm{W}^-}\xspace}
\newcommand{\Wplus}    {\ensuremath{\rm{W}^+}\xspace}
\newcommand{\Z}        {\ensuremath{\rm{Z}^0}\xspace}

\newcommand{\fivenn}   {$\sqrt{s_{\mathrm{NN}}} = 5.02$~Te\kern-.1emV\xspace}
\newcommand{\eightnn}  {$\sqrt{s_{\mathrm{NN}}} = 8.16$~Te\kern-.1emV\xspace}
\newcommand{\thirteen} {$\sqrt{s} = 13$~Te\kern-.1emV\xspace}
\newcommand{\GeVc}     {Ge\kern-.1emV$/c$\xspace}
\newcommand{\GeVmass}  {Ge\kern-.2emV$/c^2$\xspace}

\newcommand{\pt}       {\ensuremath{p_{\rm T}}\xspace}
\newcommand{\avTaa}    {\ensuremath{\langle T_{\rm AA} \rangle}\xspace}

\begin{document}

\title{Electroweak-boson production from small to large systems with ALICE at the LHC}
\author{\firstname{Guillaume} \lastname{Taillepied}\inst{1}\fnsep\thanks{\email{g.taillepied@cern.ch}} for the ALICE Collaboration}
\institute{Research Division and ExtreMe Matter Institute EMMI,\\ GSI Helmholtzzentrum für Schwerionenforschung GmbH, Darmstadt, Germany}


\abstract{
Recent measurements of the electroweak-boson production performed by the ALICE Collaboration are reported. The \W-boson production was measured at midrapidity in pp collisions at \thirteen, via the electronic decay of the boson. It was also measured in association with a hadron emitted back-to-back with respect to the electron. In heavy-ion collisions, the \W- and \Z-boson production was measured at forward rapidity in p--Pb collisions at \eightnn and Pb--Pb collisions at \fivenn, in their muonic decay channel. Measurements in pp collisions shed light on multiple parton interactions and colour reconnection mechanisms. In heavy-ion collisions the measurements of electroweak bosons allow to probe the initial state of the collision.
}

\maketitle

\section{Introduction}
\label{sec:intro}

One of the important observations in the LHC era is that the self-normalised production of hadrons increases faster than linearly as a function of the self-normalised charged-particle multiplicity \cite{aliceHFhadrons}. This enhancement in high-multiplicity events is not fully understood, but calculations including multiple parton interactions (MPI) and colour reconnection (CR) are able to reproduce the trend. The measurement of the multiplicity-dependent production of the \W boson, which is colourless, with associated hadron in proton--proton (pp) collisions can then provide further understanding of the MPI and CR mechanisms.

In heavy-ion collisions (HIC), electroweak bosons are valuable probes of the initial phase of the collision, on which a precise knowledge is required to disentangle QGP-induced phenomena from other nuclear effects. Being hard processes, the production of the \W and \Z bosons is highly sensitive to the initial state, especially the modifications of the Parton Distribution Functions (PDF) in the nucleus with respect to that in the proton. The bosons decay before the typical time of creation of the QGP and can be measured via their leptonic decays, providing a final state insensitive to the strong force. The whole process is then medium-blind, carrying the information from the initial state to the detector where it can be recorded. \\

The measurements presented here were performed using data collected with the ALICE detector~\cite{alice}. In pp collisions at \thirteen, electrons from \W-boson decays were measured at midrapidity, in the interval $|y| < 0.6$. Electrons coming from \W-boson decays typically have a large transverse momentum (\pt) and are well isolated from other particles. They were thus identified by considering electrons in the range $30 < \pt^{\rm e} < 60$~\GeVc, then defining an isolation cone surrounding the electron in which the total energy is requested to be below a certain threshold. The associated hadrons, produced together with \W bosons, were detected away-side in azimuth with respect to the electron coming from the decay of the boson, applying a lower minimum \pt selection of 5~\GeVc.

Measurements in HIC were performed in proton--lead (p--Pb) collisions at \eightnn and lead--lead (Pb--Pb) collisions at \fivenn. The data were collected from muon-triggered events in the muon forward spectrometer, covering the rapidity interval $2.5 < y < 4$. The \Z-boson candidates were reconstructed by combining high-\pt muons ($\pt^\mu > 20$ \GeVc) in pairs of opposite charge, considering the pairs with an invariant mass in the range $60 < m_{\mu^+\mu^-} < 120$ \GeVmass. The number of \Wminus and \Wplus bosons was extracted via a template fit to the single-muon \pt distribution, accounting for the various contributions to the inclusive spectrum. As the low-\pt region features a very low signal-to-background ratio, the measurements were performed on muons with \pt > 10 \GeVc.

The available measurements of electroweak bosons performed by the ALICE Collaboration at forward rapidity can be found in Refs.~\cite{aliceWZ,aliceZ,aliceW}.

\section{Results}
\label{sec:results}

\subsection{Measurements in pp collisions}
\label{sec:pp}

\begin{figure}[h]
    \centering
    \sidecaption
    \includegraphics[width=0.5\linewidth,clip]{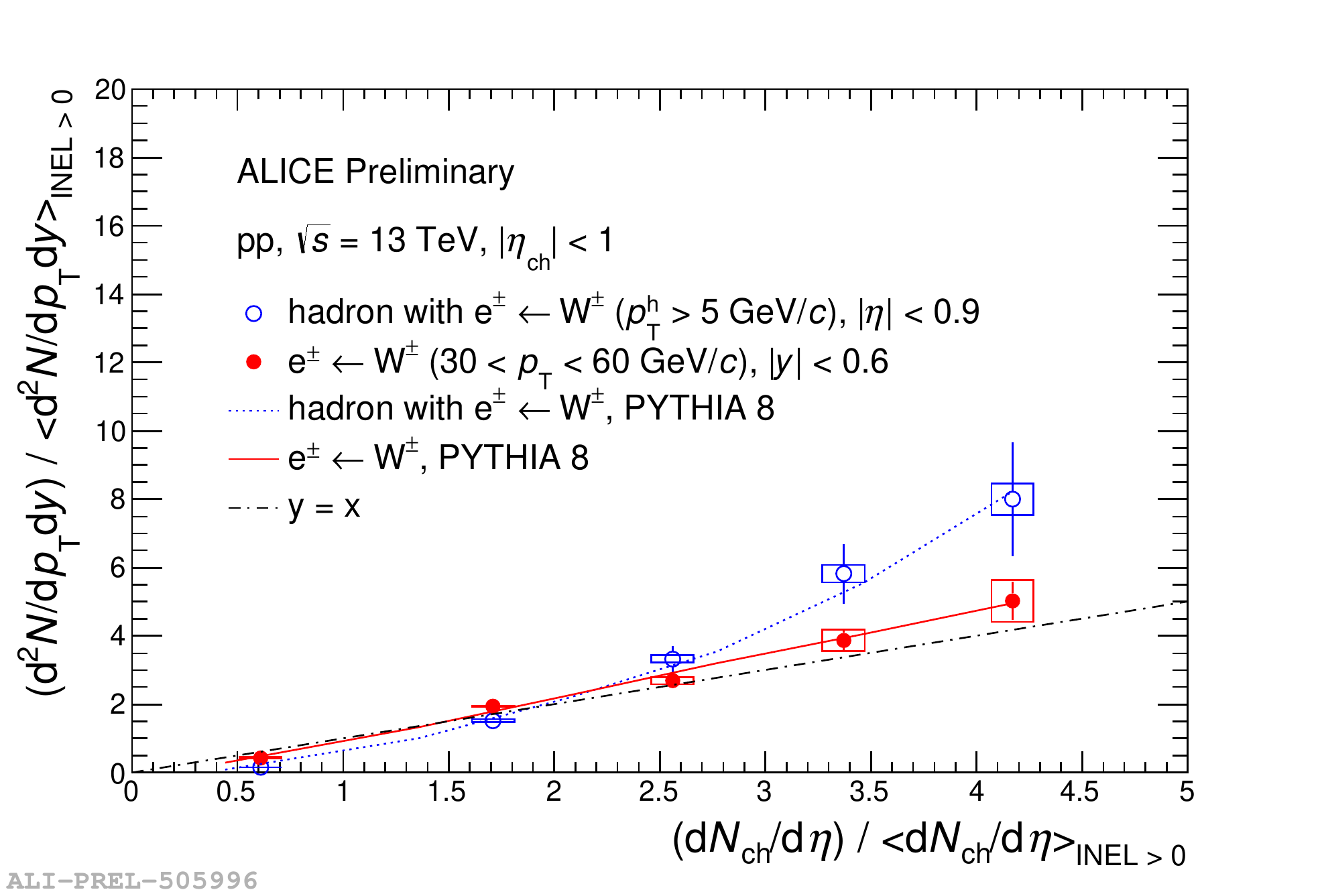}
    \caption{Self-normalised yield of electron from \W decays (red), and hadron associated with \W bosons (blue), as a function of the charged-particle multiplicity. The coloured line correspond to calculations with PYTHIA~8~\cite{pythia8} including MPI and CR.}
    \label{fig:pp}
\end{figure}

Figure~\ref{fig:pp} shows the self-normalised multiplicity-dependent yield of electrons from \W boson decays (combining \Wminus and \Wplus for increased precision), and that of hadrons associated with a \W boson, measured in pp collisions at \thirteen. The yield of electrons from \W-boson decays is consistent with a linear increase as a function of the charged-particle multiplicity, while the yield of associated hadron shows a faster-than-linear increase. This measurement thus suggests a significant correlation of the associated hadron production with the event multiplicity, and the trend is well reproduced by PYTHIA~8~\cite{pythia8} calculations including MPI and CR mechanisms. This analysis also constitutes the first measurement of electroweak bosons at midrapidity with ALICE, and will serve as reference for similar measurements in HIC.

\subsection{Measurements in p--Pb collisions}
\label{sec:ppb}

\begin{figure}[h]
    \centering
    \includegraphics[width=0.43\linewidth,clip]{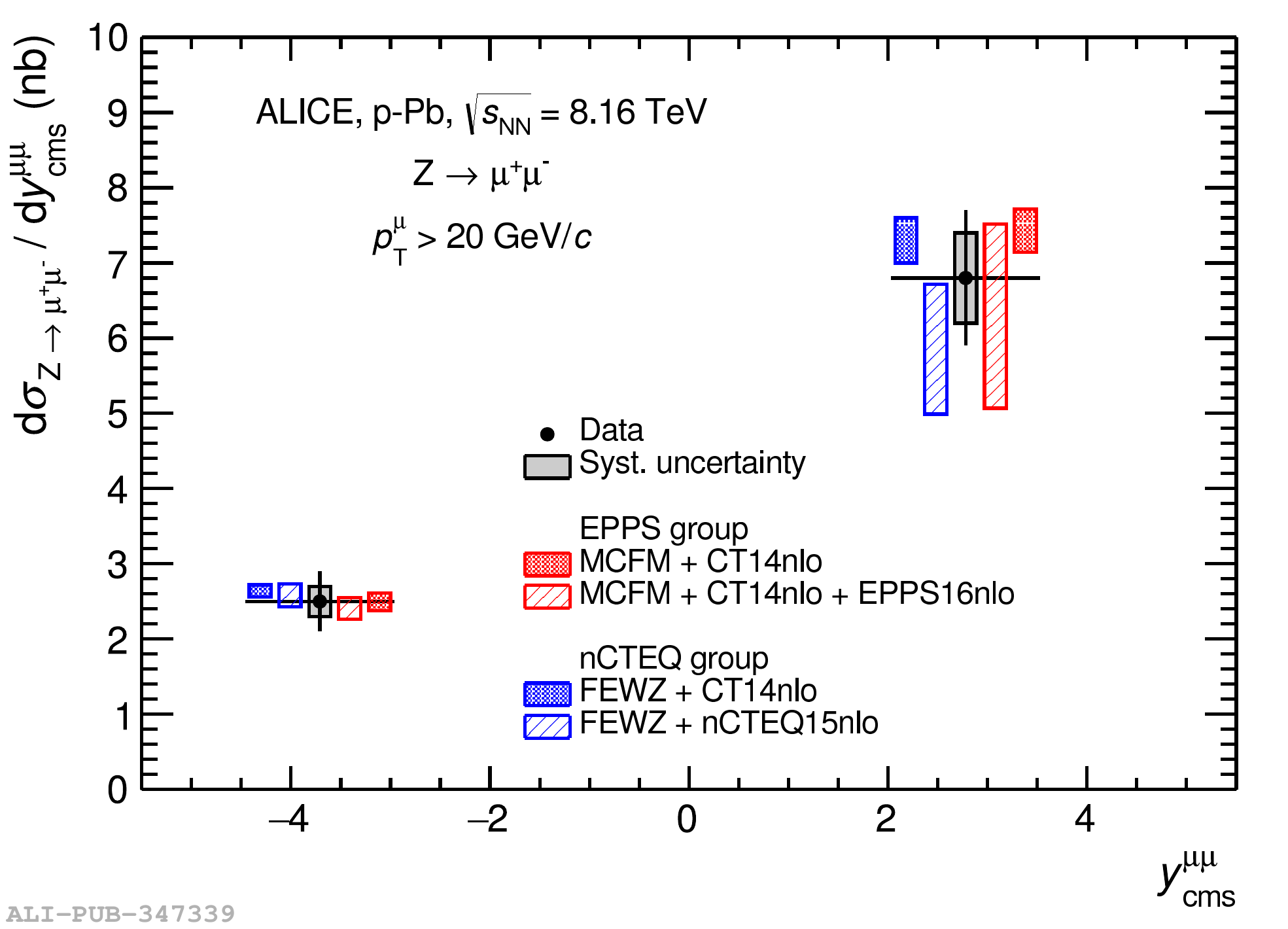}
    \includegraphics[width=0.55\linewidth,clip]{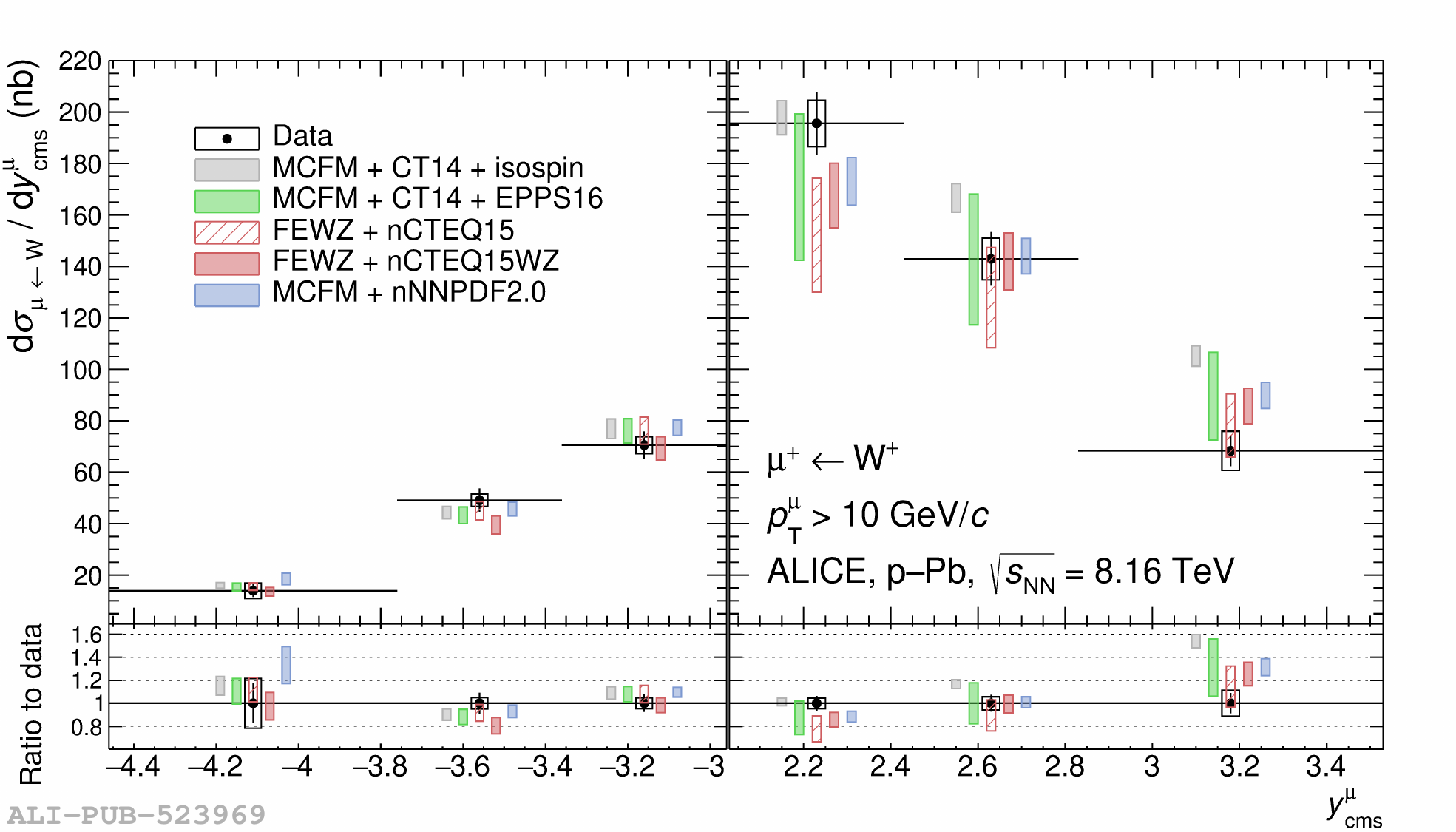}
    \caption{\textbf{Left}: cross section of dimuons from \Z decays measured in p--Pb collisions at \eightnn, compared with theoretical calculations with and without nuclear modifications of the PDF. \textbf{Right}: cross section of muons from \Wplus decays in the same collisions systems, compared with model predictions from various PDF and nPDF sets.}
    \label{fig:ppb}
\end{figure}

The measurements of the \Z and \Wplus bosons as a function of rapidity in p--Pb collisions at \eightnn are shown in the left and right panels of Fig.~\ref{fig:ppb}, respectively. The larger number of events available for the \Wplus boson allowed splitting the acceptance into several rapidity intervals. The \Z-boson measurement is compared with predictions from the EPPS16~\cite{epps16} and nCTEQ15~\cite{ncteq15} nuclear PDFs (nPDFs), as well as predictions from the CT14~\cite{ct14} free PDF accounting for the isospin effect but without nuclear modifications. The models are in good agreement with the measurement, although no strong conclusion can be derived on the nuclear modifications due to statistical limitations.

The \Wplus cross section shown in the right panel of Fig.~\ref{fig:ppb} is compared with the same models, as well as two recent nPDF sets: nCTEQ15WZ~\cite{ncteq15wz}, an improvement of nCTEQ15 with the addition of electroweak-boson measurements from the LHC for the nPDF determination; and nNNPDF2.0~\cite{nnnpdf2}, a new model obtained following a methodology based on machine learning. The models including nuclear modifications are in agreement with one another, and provide a good description of the data. Calculations without nuclear modifications, on the contrary, overestimate the production at large positive rapidity. The resulting 3.5$\sigma$ deviation is the strongest observation of nuclear modification of the PDF measured by the ALICE Collaboration from electroweak-boson measurements. This measurement helps constraining the nPDF models in the very low Bjorken-$x$ region (down to about 10$^{-4}$), where the available constraints are scarce. \\

\begin{figure}[h]
    \centering
    \includegraphics[width=0.46\linewidth,clip]{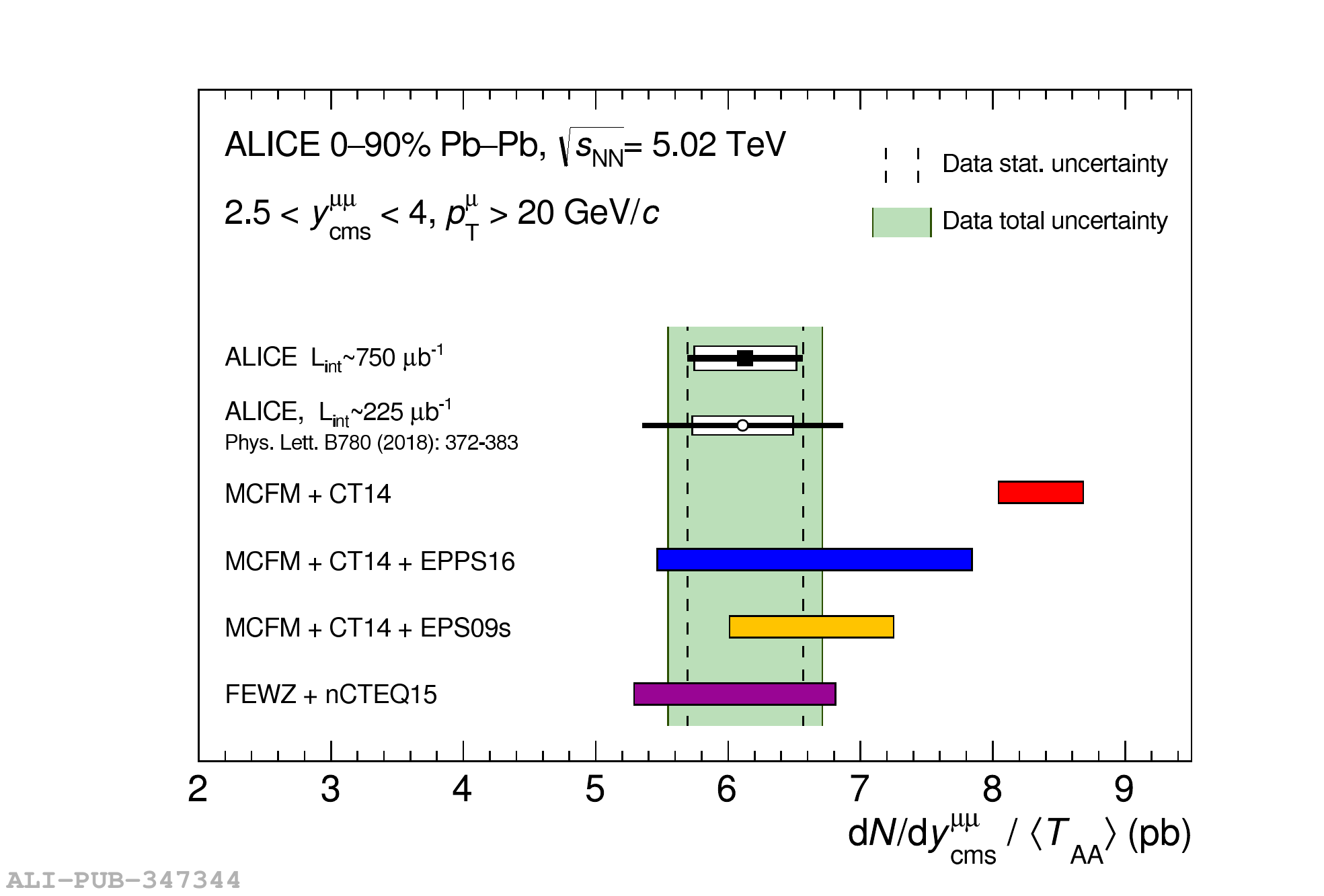}
    \includegraphics[width=0.52\linewidth,clip]{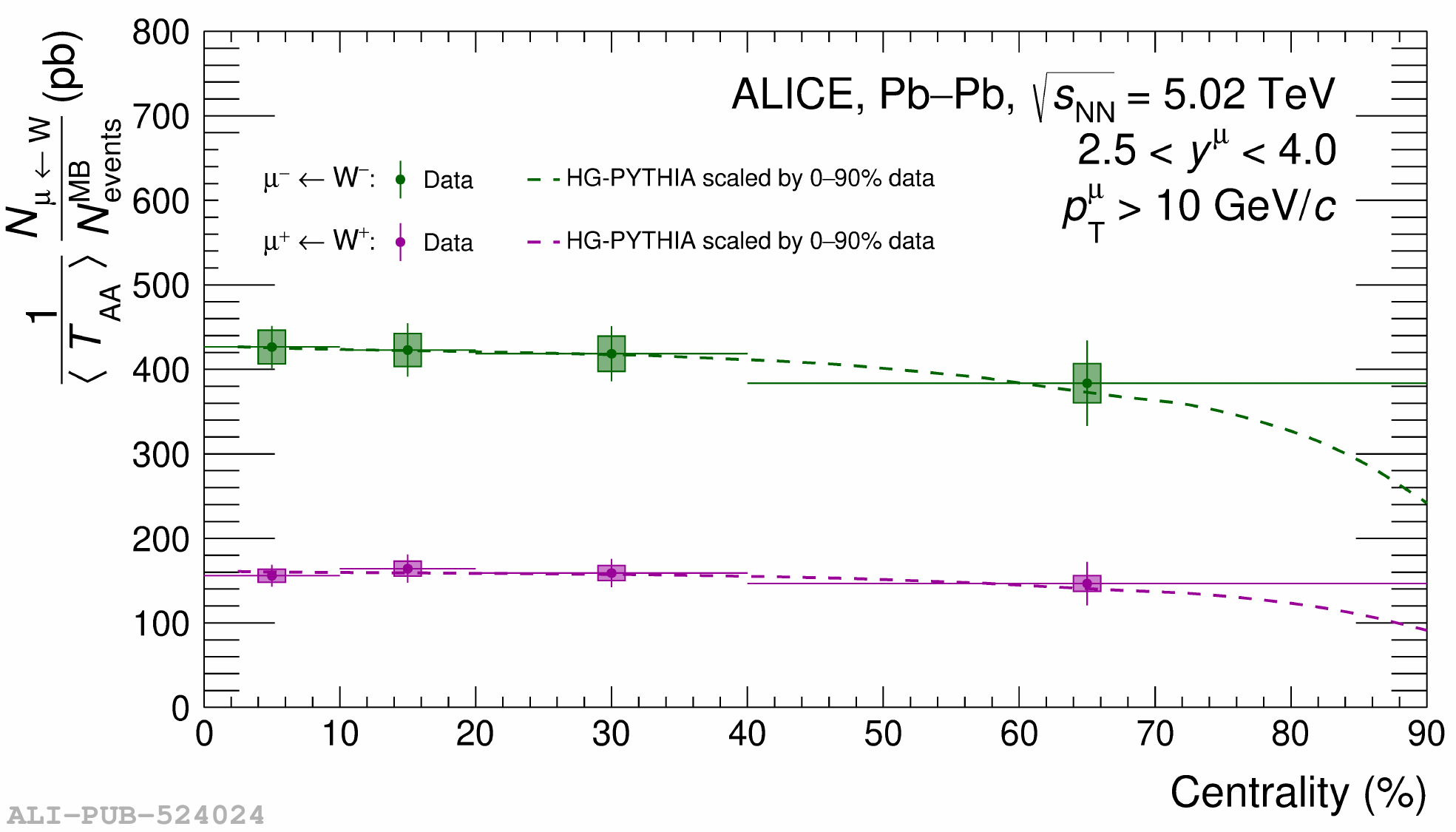}
    \caption{\textbf{Left}: \avTaa-scaled yield of dimuons from \Z-boson decays in Pb--Pb collisions at \fivenn, compared with theoretical calculations with and without nuclear modifications of the PDF. \textbf{Right}: \avTaa-scaled yield of muons from \W-boson decays as a function of centrality, in Pb--Pb at \fivenn. The measured distribution is compared with HG-PYTHIA~\cite{hg-pythia} predictions of the nuclear modification factor of hard scatterings, scaled to the measured value in 0--90\% centrality.}
    \label{fig:pbpb}
\end{figure}

\subsection{Measurements in Pb--Pb collisions}
\label{sec:pbpb}

The results obtained in Pb--Pb collisions at \fivenn are illustrated in Fig.~\ref{fig:pbpb}. In the left panel, the yield of dimuons from \Z decays, scaled with the average nuclear overlap function \avTaa, is compared with a previously published result obtained from the 2015 data only, as well as predictions with and without nuclear modifications. The combination of the 2015 and 2018 data periods for this new measurement improved the available integrated luminosity by a factor three, allowing for a significant reduction of the uncertainty. The various nPDFs provide a good description of the measured yields, while a deviation by 3.4$\sigma$ from the CT14-only calculation is observed, constituting another strong sign of nuclear modifications.

In the right panel of Fig.~\ref{fig:pbpb}, the \avTaa-scaled yield of electrons and positrons from \W decays are shown as a function of the collision centrality. They are compared with HG-PYTHIA~\cite{hg-pythia} calculations of the nuclear modification factor of hard scatterings, scaled with the value measured in 0--90\% centrality. The scaled calculations are in good agreement with the data. They predict a sizeable drop for the most peripheral collisions, but statistical limitations prevent to have enough granularity in this region in the measurement, such that no experimental conclusion can be drawn. Nonetheless, this measurement is the first measurement of the \W-boson production in Pb--Pb collisions at forward rapidity, where low values of Bjorken-$x$ are attained.

\section{Conclusion}

The recent measurements of electroweak-boson production show their usefulness in the characterization of key features of QCD. The measurement at midrapidity in pp collisions is well described by calculations including MPI and CR. In heavy-ion collisions, the measurements are generally well described by nuclear models, while significant deviations from free-PDF calculations can be observed. These measurements can thus be used to help constraining the nPDF models.

\end{document}